\documentclass[aps,prb,twocolumn,groupedaddress,showpacs]{revtex4}
\usepackage{graphicx}
\usepackage{isolatin1}
\begin{document}

\title{Low-temperature behavior of two-dimensional Gaussian Ising spin
glasses}

\author{Jérôme Houdayer}
\affiliation{Service de Physique Théorique, 
CEA Saclay, 91191 Gif-sur-Yvette, France}

\author{Alexander K. Hartmann}
\affiliation{Institut für Theoretische Physik, Universität Göttingen,
Tammannstra\ss{}e 1, 37077 Göttingen, Germany}

\date{\today}

\begin{abstract}
We perform Monte Carlo simulations of large two-dimensional Gaussian
Ising spin glasses down to very low temperatures
$\beta=1/T=50$. Equilibration is ensured by using a cluster algorithm
including Monte Carlo moves consisting of flipping  fundamental 
excitations. We study the thermodynamic behavior using the Binder
cumulant, the spin-glass susceptibility, the distribution of overlaps,
the overlap with the ground
state and the specific heat. We confirm that $T_c=0$. All results are
compatible with an algebraic divergence of the correlation length with
an exponent $\nu$. We find $-1/\nu=-0.295(30)$, which is compatible with
the value for the domain-wall and droplet exponent
$\theta\approx-0.29$ found previously in ground-state studies. Hence
the thermodynamic behavior of this model seems to be
 governed by one single exponent.

\end{abstract}

\pacs{75.50.Lk, 05.70.Jk, 75.40.Mg, 77.80.Bh}

\maketitle

\section{Introduction}

Spin glasses \cite{reviewSG} are the prototype models for disordered
systems investigated extensively  in statistical physics 
during the last three decades.  These systems exhibit complex energy landscapes
resulting in many interesting phenomena, like glassy behavior and
aging.  Despite much effort, many questions are still open.

For two-dimensional spin glasses with only nearest-neighbor
interactions, it is now clear that no stable spin-glass phase at
finite temperature exists
\cite{rieger1996,kawashima1997,stiff2d,houdayer2001,carter2002}.
This means that they are paramagnetic at any finite temperature and
that they exhibit spin-glass behavior only at $T=0$.  Thus, it is
widely believed, that their low-temperature behavior can well be
described by the droplet theory \cite{McM,BM,FH}.  The droplet picture
assumes that the low-temperature behavior is governed by droplet-like
excitations, where excitations of linear spatial extent $l$ typically
cost an energy of order $l^{\theta}$. Thus in the thermodynamic limit
the excitations which flip a finite fraction of the spins cost an
infinite amount of energy if $\theta>0$.  For the two-dimensional spin
glass with Gaussian interactions, since it exhibits no order at $T>0$,
$\theta<0$ holds \cite{comment_pmJ}.  Furthermore it is usually
assumed that the energy of different types of excitations,
e.g. droplets and domain walls, induced by changing the boundary
conditions, are described by the same exponent $\theta$. Indeed,
recently is has been confirmed by calculating exact ground states
\cite{opt-phys2001} that $\theta=-0.282(2)$ for droplet and
domain-wall excitations
\cite{aspect-ratio,droplet2d}. For domain walls, small sizes are
sufficient to see the asymptotics, hence similar values have been
found previously
\cite{rieger1996,bray1984,mcmillan1984,mcmillan1984b,cieplak1990,%
matsubara1998,KA} in this case. On the other hand, for some droplet
excitations, this behavior is only visible for not-too small system
sizes $L\ge 50$, which explains why in a similar preceding study
\cite{kawashima2000} of smaller sizes an apparently different exponent
close to $\theta=-0.47$ has been found. For other types of
droplet-like excitations, the exponent $\theta\approx -0.28$ is again
already visible for small sizes \cite{excited2d,berthier2003}.  Hence,
the behavior at zero temperature seems to be relatively clear
\cite{remark_picco}.

The situation is different for the small but finite-temperature behavior.
In case the correlation length $\xi$ diverges algebraically for $T\to 0$
like $\xi\sim T^{-\nu}$, the critical exponent $\nu$ of the
correlation length is related through a simple renormalization
argument \cite{bray1984,mcmillan1984} to the droplet exponent $\theta$ via
$\theta=-1/\nu$. Several numerical studies to obtain $\nu$ at finite
temperatures have been performed. Using transfer-matrix calculations
of long  ($L_x$ up to $10^6$) and narrow ($L_y$ up to 11) stripes, 
values of \cite{cheung1983b} $\nu=2.96(22)$  respectively \cite{huse1985} 
$\nu=4.2(5)$ have been found. For small ($L=10$) square systems 
\cite{kawashima1992}, a value of $\nu=2.1(1)$ was found.
Also a couple of Monte Carlo (MC) simulations have been performed. For
small sizes ($L=12$) and relative large temperatures $T\ge 1$, a value
of $\nu=3.6(2)$ has been found \cite{rieger1996}. Later, for similar
system sizes but lower temperatures, a value of $\nu=1.8(4)$ was
obtained \cite{ney-nifle1997}.  Furthermore, a cluster Monte Carlo
simulation of three large samples ($L=128$) in the temperature range
$T\in [0.4,1]$ has bee performed, resulting
\cite{liang1992} in $\nu=2.0(2)$.

Since the finite-temperature studies performed so far suffer from
either too small systems or too large temperatures (or both), we have
performed a Monte-Carlo study of large systems up to size $L=75$. By
applying a recently developed cluster algorithm \cite{houdayer2001},
in connection with an extension presented below, we are able to
equilibrate the system at much smaller temperatures than it was
possible before. Parallel and independently of us, H.G. Katzgraber,
L.W. Lee and A.P. Young performed a related study
\cite{katzgraber2004} using a similar algorithm
. 
They focus on the direct calculation of the
correlation length, while we study here other thermodynamic quantities like 
the Binder parameter, the spin-glass susceptibility, the distribution
of overlaps, the overlap with the ground state and the specific heat, 
and we infer from these results the asymptotic behavior of the
correlation length indirectly.

The model we study consists of $N=L^2$\ Ising spins $S_i=\pm 1$ on a
square lattice with the Hamiltonian
\begin{equation}
{\cal H} = -\sum_{\langle i,j \rangle} J_{ij} S_iS_j\,,
\end{equation}
where the sum runs over all pairs of nearest neighbors $\langle i,j
\rangle$ and the $J_{ij}$ are quenched random variables which are
distributed according to a Gaussian distribution with zero mean and
unit variance. In the following, we denote the thermal average by
$\langle\ldots\rangle$ and the quenched-disorder average by
$[\ldots]_J$. Periodic boundary conditions in both directions are
applied.

The rest of the paper is organized as follows. First we explain the
algorithm we have applied and show that it is able to equilibrate the
system. In the main part, we present our results for the different
thermodynamic quantities mentioned above. In the last
section, we summarize our results.

\section{Algorithm}

We have made extensive MC simulations of our system. To reach
equilibrium down to very low temperatures for large system sizes, 
we have used a recently developed cluster algorithm 
(details can be found  elsewhere \cite{houdayer2001}). 
To speed up equilibration at very low temperatures 
we have devised and used a new
procedure. It consists in maintaining a list of the lowest-energy
elementary excitations, and to use flipping them as MC moves. This
procedure works as follows:

\begin{enumerate}
\item Compute the ground state of the system. To do this we used a
heuristic renormalization-group based algorithm \cite{houdayermartin99,
houdayermartin01}. To test the efficiency of the method, we checked, 
for systems of size $L=100$ with open boundaries, 
that this algorithm systematically finds the true ground state by
comparing with the result of an exact algorithm 
\cite{opt-phys2001}  (which works only for planar
graphs, i.e. not with
fully periodic boundary conditions). Moreover during the production runs
using the MC 
simulations, we never found  excitations with negative energy, which
confirms that the true ground state has been found.

\item During the equilibration phase, we systematically compare the
low temperature ($T\le 0.2$) configurations with the ground
state. They differ by clusters that are flipped. We maintain a list of
the lowest-energy excitations thus found (we keep up to 10000 of such
excitations for the largest system). Note that we consider only {\it
elementary} excitations whose boundary is connected (two clusters
flipped inside one another define two independent elementary excitations, not
one). Then we introduce (in addition to the cluster algorithm) a new
Monte Carlo move at low temperature ($T \le 0.2$): choose an elementary
excitation in the list and try to flip it using the Metropolis
criterion. As soon as the list no longer evolves ({\it i.e.} when all
the first excitations have been found) these moves trivially respect
detailed balance.

\item During the production phase, we no longer try to find new
excitations, we simply use the cluster algorithm together with 
single-spin flips.
\end{enumerate}

Note that we obtain a list of the first elementary excitations as a
by-product of this algorithm. To check that equilibrium has been
reached, we have used the criterion described in Ref.\
\onlinecite{katzgraber2001}, which is based on the following identity
valid for a Gaussian distribution of the interactions:

\begin{equation}
[E]_J = - \beta N_l (1 - [\langle q_l\rangle]_J)
\label{eqEquilibration}
\end{equation}
where $E=\langle {\cal H} \rangle$ is the average energy, $\beta=1/T$ the
inverse temperature, $N_l=2N$ is the 
number of bonds, and $q_l$ denotes the link overlap between two independently
chosen configurations $\{S_i^\alpha\}$, $\{S_i^\beta\}$ for the same disorder:

\begin{equation}
q_l = \frac{1}{N_l} \sum_{\langle i, j\rangle} S_i^\alpha S_j^\alpha
S_i^\beta S_j^\beta.
\end{equation}

\begin{figure}
\begin{center}
\resizebox{0.9\linewidth}{!}{\includegraphics{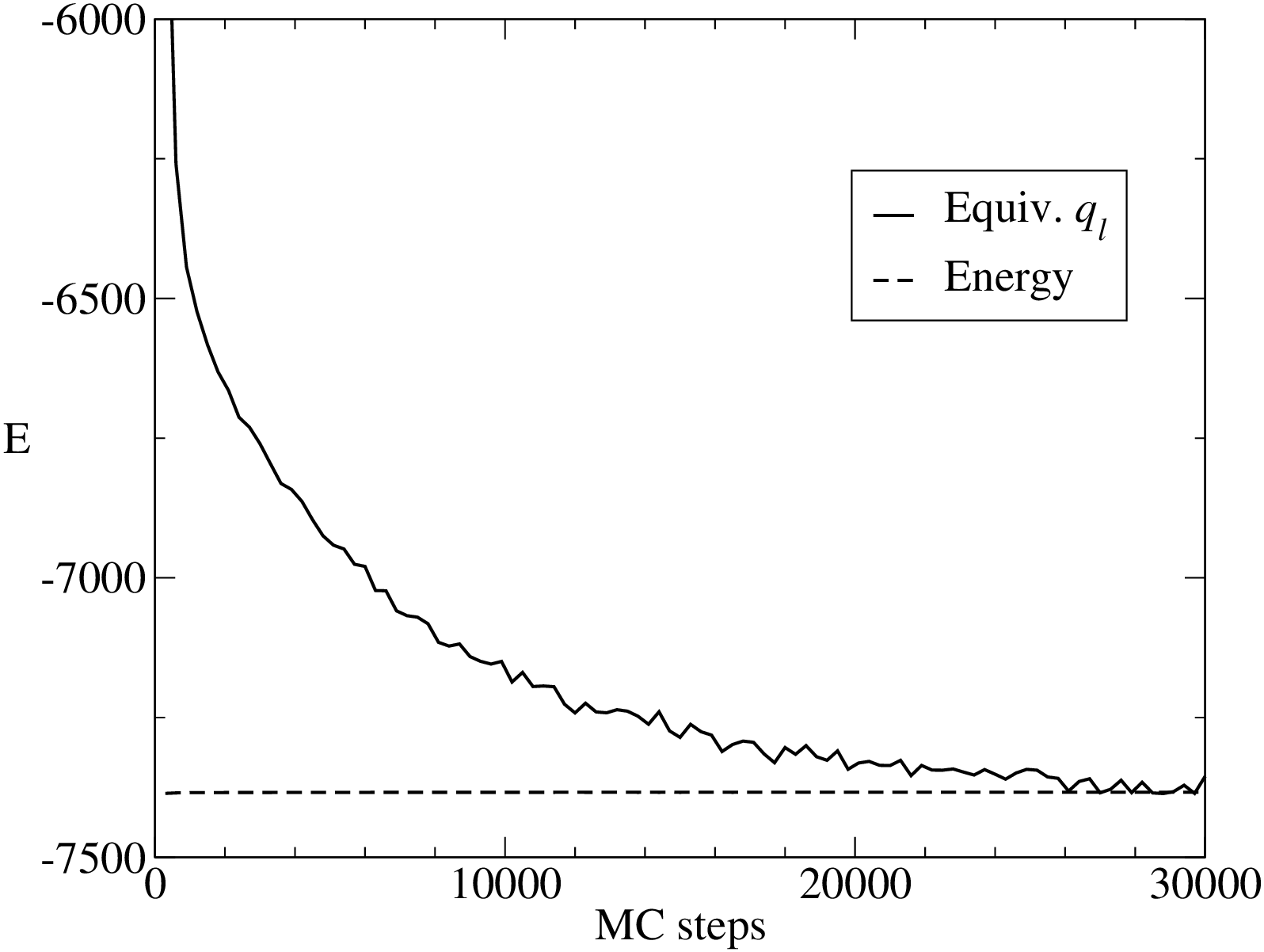}}
\end{center}
\caption{Sample equilibration test for $L=75$ and $\beta=5$ averaged over
200 samples, the curves (Eq.\ \ref{eqEquilibration}) start to overlap
around the equilibration time (here 25000 MC steps).}
\label{figEquilibration}
\end{figure}

In Fig.\ \ref{figEquilibration} we show how both sides of
Eq. \ref{eqEquilibration} evolve during equilibration. The system
can be considered equilibrated when the curves start to overlap. 
Note that within our algorithm, the
temperature, where equilibration takes in longest time, 
is not the lowest temperature, because the
excitation flips are most efficient at the lowest temperatures.
 For example at the lowest temperature, the
ground state may be found after the very first step of the algorithm
because all excitations present in the starting configurations can be
flipped at once.

\section{Results}

We considered the following sizes for our simulations: 
$L=10$, 25, 35, 50 and 75 over a very large
range of temperatures $0.02\le T\le 5$. We respectively treated 1000,
1000, 1000, 500, 200 samples for the different sizes. 
We simulated at different values of the temperature, the number of different
values ranging from 19 to 59 (with increasing size). For each sample,
we simulated 64 independent configurations at each temperature.

\begin{figure}
\begin{center}
\resizebox{0.9\linewidth}{!}{\includegraphics{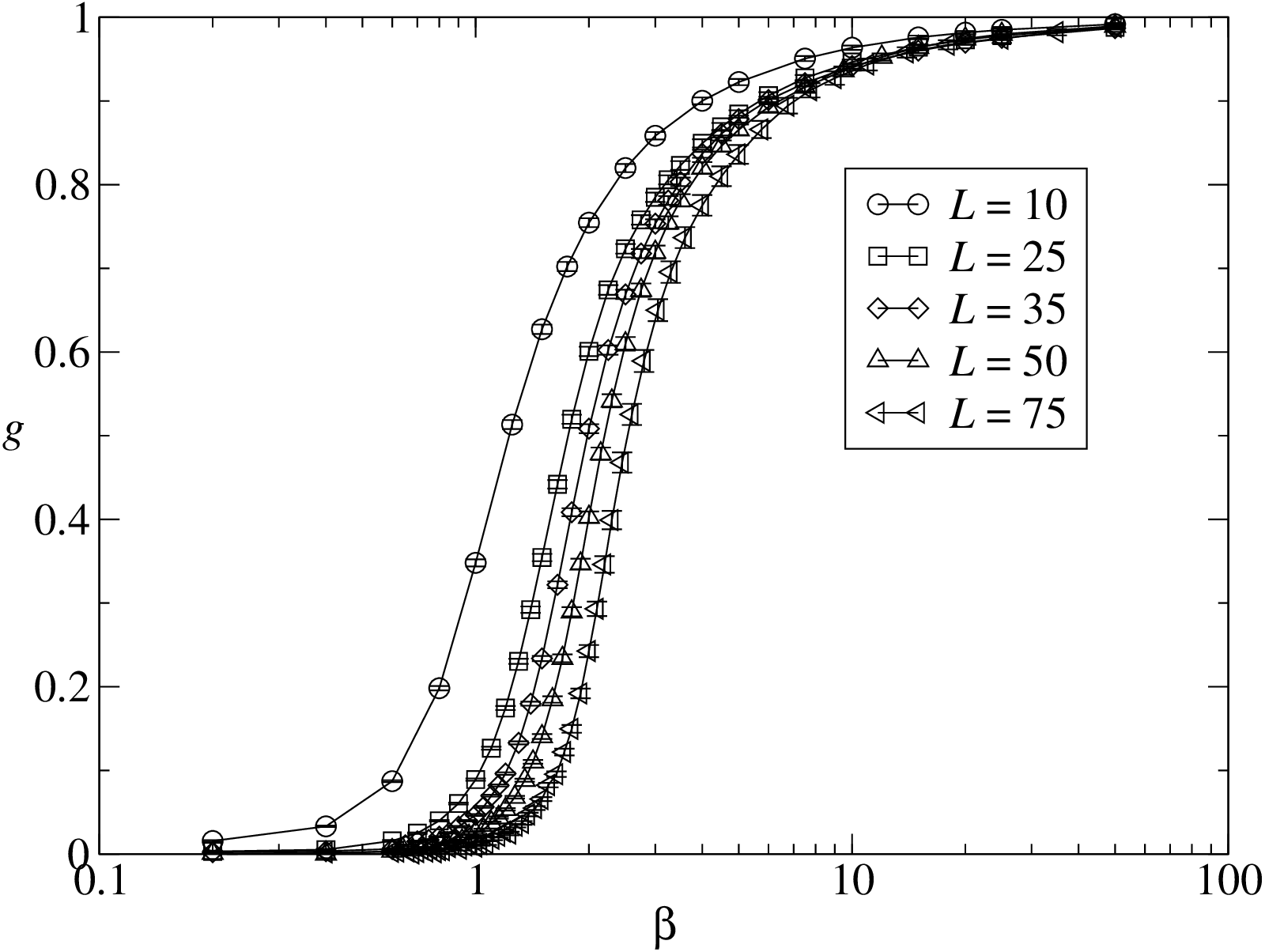}}
\end{center}
\caption{Binder cumulant $g$ as a function of inverse temperature
  $\beta$ for different system sizes L.}
\label{figRawBinder}
\end{figure}

To study our system, we have measured different quantities and averaged
over the different samples. Our first quantity of interest is the
Binder cumulant \cite{binder81,bhatt85} $g$ defined by
\begin{equation}
g = \frac12 \left[3-\frac{\langle q^4\rangle}{\langle q^2\rangle^2}\right]_J.
\end{equation}
Here $q$ is the overlap between two independent 
equilibrated configurations $\{S_i^\alpha\}$ and $\{S_i^\beta\}$ of
the same disorder realization
\begin{equation}
q = \frac1N\sum_i S_i^\alpha S_i^\beta.
\end{equation}
  In the thermodynamic limit,
the Binder cumulant is zero in the paramagnetic phase and around one in the
spin-glass phase. To study $g$ at finite sizes, we consider the
divergence of the correlation length $\xi$ when approaching the
transition temperature $T_c$
\begin{equation}
\xi \sim (T-T_c)^{-\nu}.
\end{equation}
Since $g$ does not show any critical behavior near the critical point,
and according to the basic assumption that it is a function of the
relation of system size $L$ to the correlation length $\xi$,  $g$  scales as:
\begin{equation} 
g \sim \tilde{g}(L/\xi) = \tilde{g}(L (T-T_c)^\nu).
\label{eq:binder-scale}
\end{equation}
Hence, $g$ is independent of $L$ at $T=T_c$ which allows the location of
$T_c$: the curves for different system sizes $L$ should intersect at
$T_c$.  In Fig.\ \ref{figRawBinder} we show the value of $g$ as a
function of $\beta$ for different system sizes $L$. 
The curves for $g$ converge to a
crossing point at very low temperature which indicate that we most
probably have $T_c=0$, in accordance to the recent believe
\cite{rieger1996,kawashima1997,stiff2d,houdayer2001,carter2002}.
 
\begin{figure}
\begin{center}
\resizebox{0.9\linewidth}{!}{\includegraphics{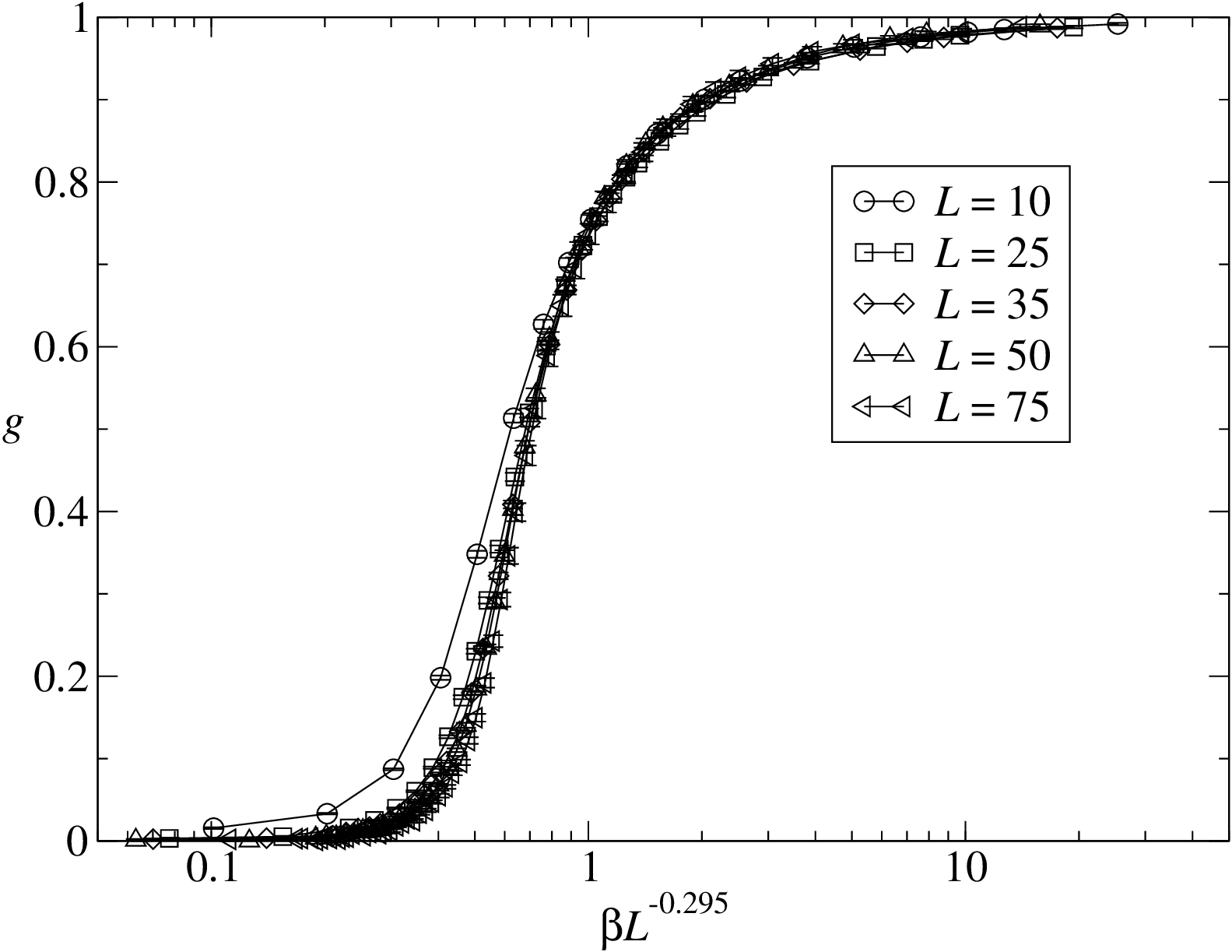}}
\end{center}
\caption{Scaling of the Binder cumulant in the critical region
  obtained by plotting $g$ as a function of the rescaled inverse
  temperature $\beta L^{-1/\nu}$ using $1/\nu=0.295$}
\label{figHighBinder}
\end{figure}

As stressed in the introduction, the value of
$\nu$ has been an open question for a long time. The reason for this
is the presence of large finite-size corrections to scaling. 
We see this, when trying to perform a finite-size scaling plot,
i.e. when plotting $g$ as a function of $\beta L^{-1/\nu}$ for a suitably
chosen value of $\nu$. 
In fact, no value for $\nu$ allows
the whole curves to collapse on a master curve. One has to select only
a domain of parameters near the critical regime, namely large $L$ and
small $T$ (and thus large $g$). Keeping only $L\ge35$ and $g\ge0.5$ we
find that $1/\nu\simeq 0.295$ ($\nu\simeq 3.39$). The scaling plot for
this value is shown on Fig.\ \ref{figHighBinder} (note that we show
all the data points, but only those in the range mentioned above 
where used to find the exponent): for large sizes and low
temperatures, a very good data collapse is obtained.

\begin{figure}
\begin{center}
\resizebox{0.9\linewidth}{!}{\includegraphics{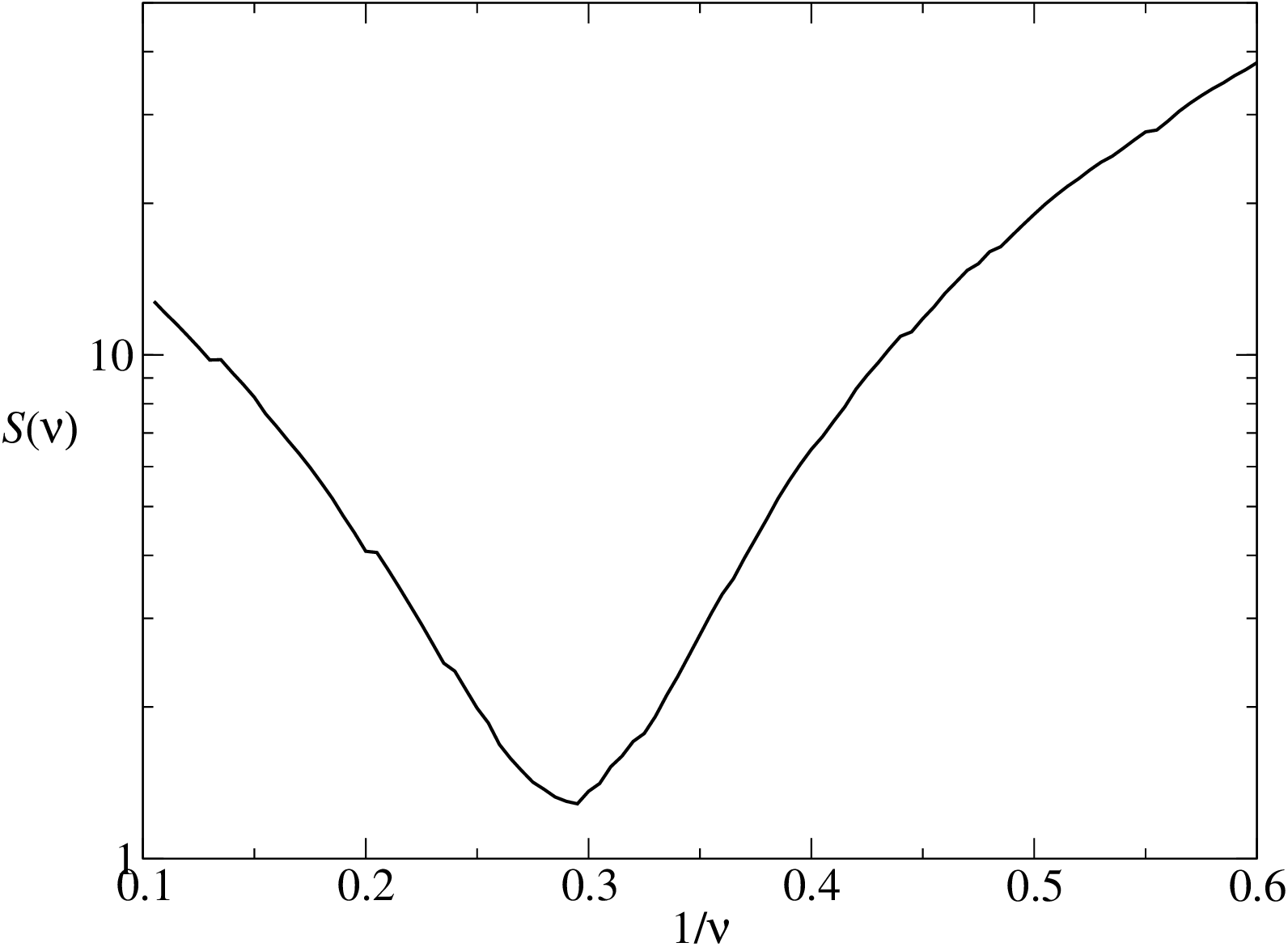}}
\end{center}
\caption{Quality function $S(\nu)$ for the Binder-cumulant scaling. The
minimum gives the estimation for the exponent $\nu$.}
\label{figS}
\end{figure}

To confirm this value of $\nu$ and to estimate the error, we need to
somehow quantify the quality of the collapse of the curves. To do
this, we use a procedure similar to one proposed by Kawashima and Ito
\cite{kawashima1993} which we detail in the appendix. Using this
procedure, we define a function $S(\nu)$, measuring the quality of the
fit as a function of the chosen value of $\nu$.  $S(\nu)$ should be
around one if the collapse is good (taking the error bars into
account) and much larger otherwise, it behaves somehow like a $\chi^2$
test. In Fig.\ \ref{figS} we show the value of $S(\nu)$ for the
scaling according to Eq. \ref{eq:binder-scale}, again we used only the
data close to $T=0$. We see that the minimum corresponds to
$S\simeq1.28$ which is good and we can also have an estimation for the
error on $\nu$: $1/\nu\simeq 0.295\pm 0.03$.

\begin{figure}
\begin{center}
\resizebox{0.9\linewidth}{!}{\includegraphics{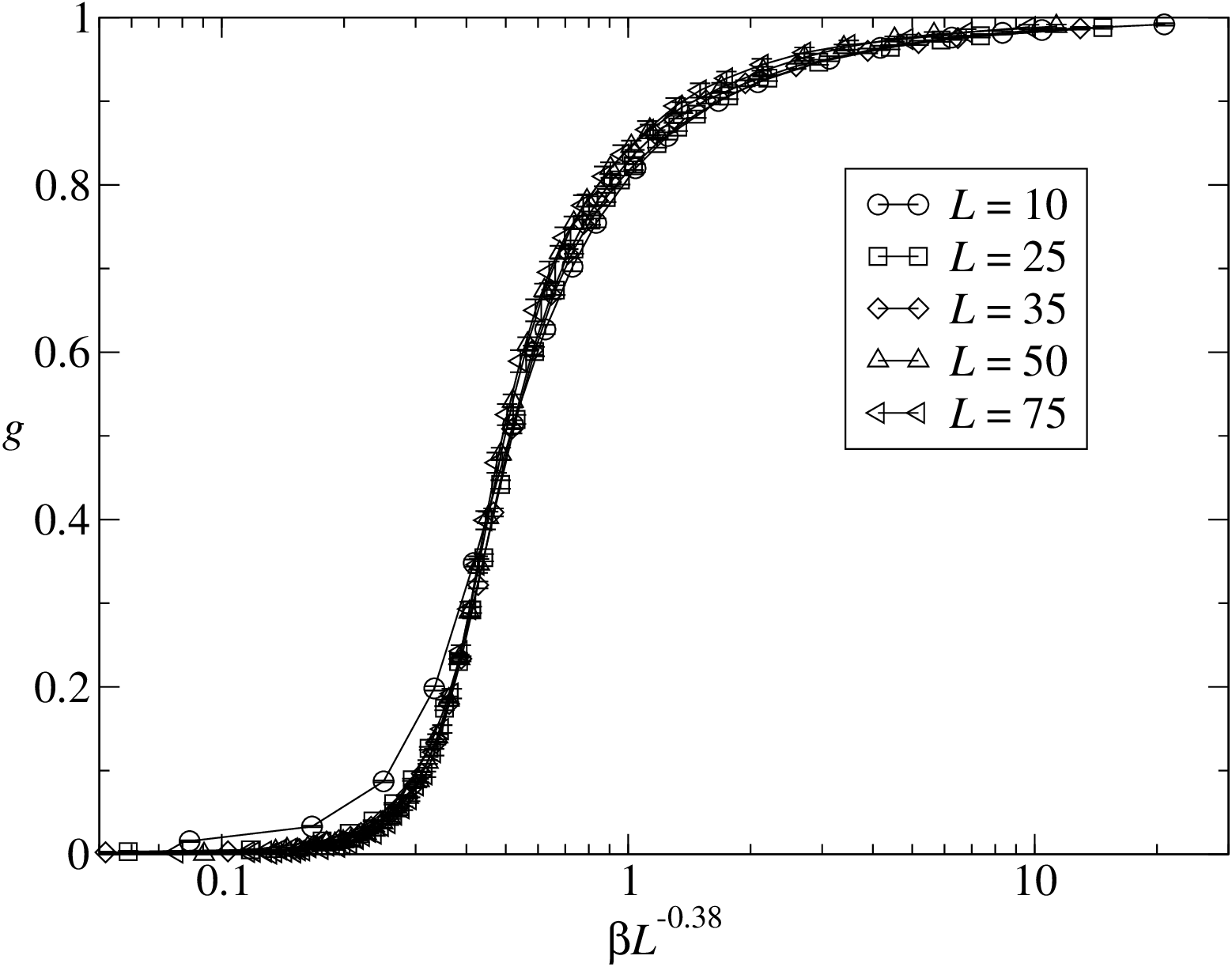}}
\end{center}
\caption{Scaling of the Binder cumulant {\it out} of the critical
  region, when choosing $1/\nu=0.38$.}
\label{figLowBinder}
\end{figure}

In Fig.\ \ref{figHighBinder}, we see that the high temperature part,
which was not used to obtain $\nu$, does not scale well. It is as if
it requires another value of the exponent. In fact, if we use only
data for which $L\ge25$ and $0.1\le g\le 0.5$ we find $1/\nu\simeq
0.38$ (and $S=6.6$ which is not good) the resulting plot is shown in
Fig.\ \ref{figLowBinder}. The scaling is not good but it explains why
such a high value of $1/\nu$ appears in previous papers: large system
sizes, low temperatures and a good criterion for equilibration are
necessary to obtain valid results.

\begin{figure}
\begin{center}
\resizebox{0.9\linewidth}{!}{\includegraphics{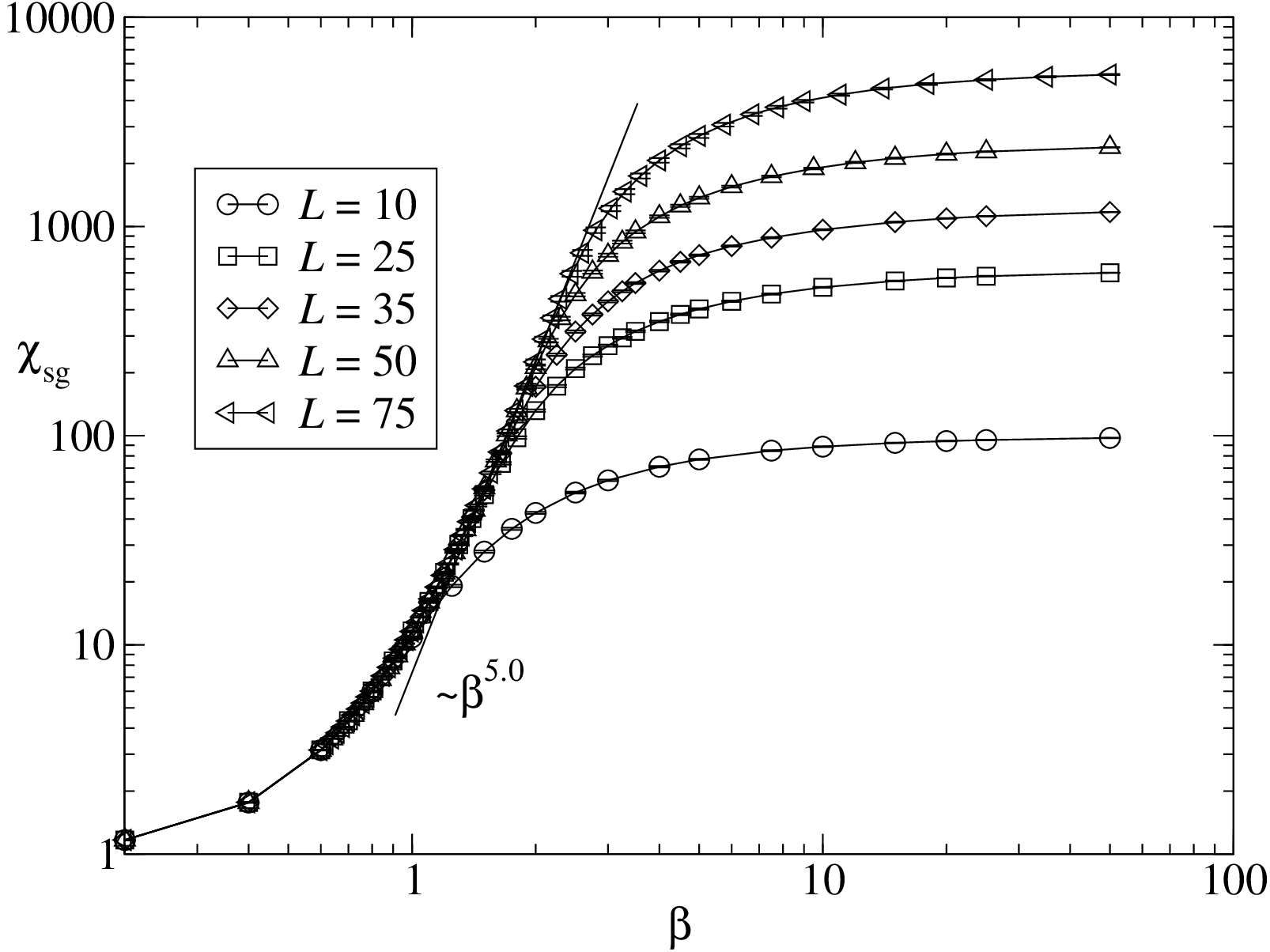}}
\end{center}
\caption{The spin-glass susceptibility $\chi_{\rm sg}$ as a function
  of inverse temperature $\beta$ for different system sizes $L$.}
\label{figChi}
\end{figure}

\begin{figure}
\begin{center}
\resizebox{0.9\linewidth}{!}{\includegraphics{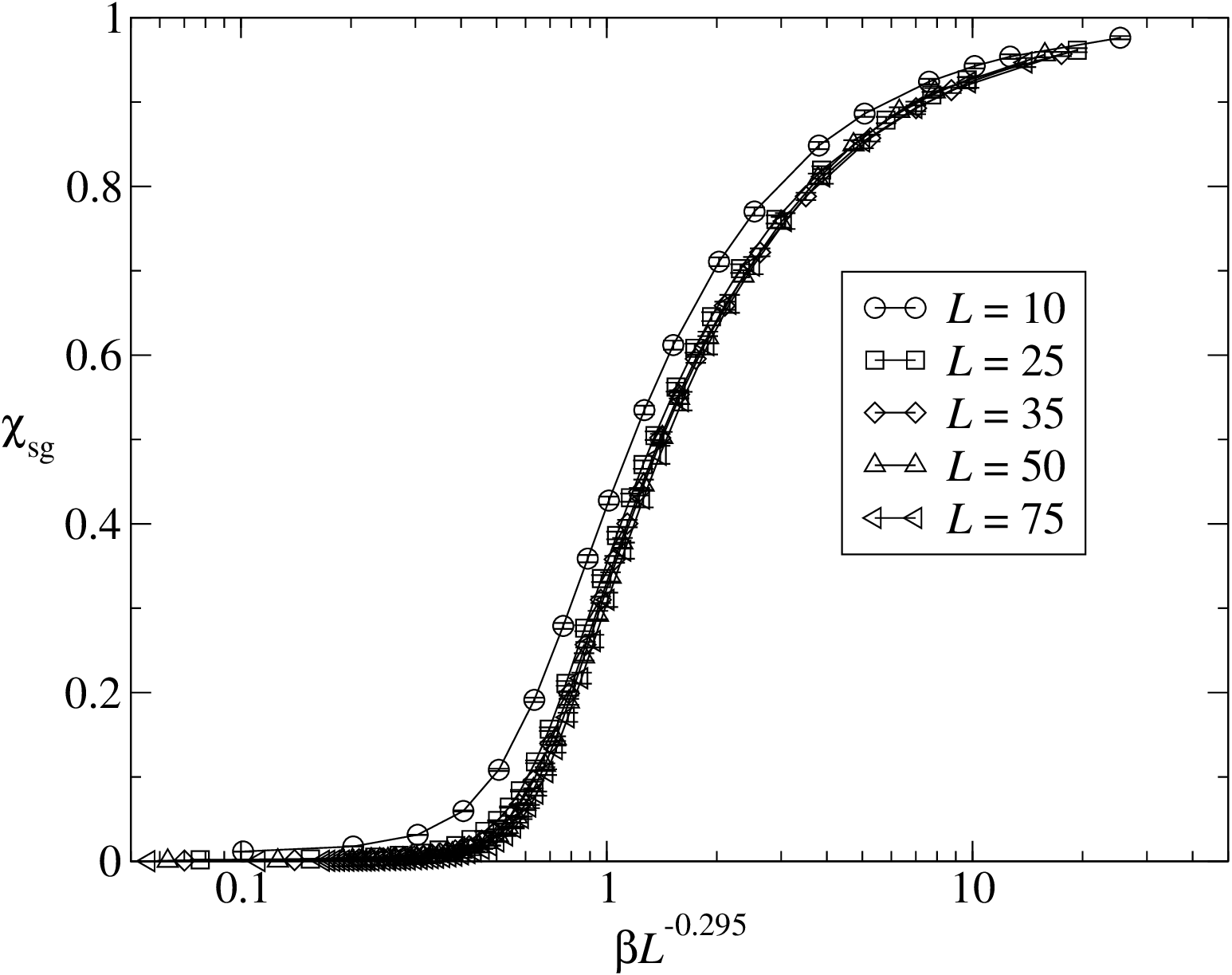}}
\end{center}
\caption{Scaling of the spin-glass susceptibility $\chi_{\rm sg}$}
\label{figScaleChi}
\end{figure}

We now turn to the spin-glass susceptibility
\begin{equation}
\chi_{\rm sg} = L^d [\langle q^2\rangle]_J
\end{equation}
(here $d=2$). Since the ground state of Ising spin
glasses with a Gaussian distribution of the interactions is unique 
(i.e.\ $q=1$), the susceptibility shows
\cite{kawashima1992,ney-nifle1997} the following finite-size behavior 
at low temperatures:
\begin{equation}
\chi_{\rm sg} \sim L^2 \tilde{\chi}(L/\xi) \sim L^2\tilde{\chi}(LT^\nu),
\label{eqScaleChi}
\end{equation}
(with $\tilde{x}(L/\xi)$ going to a constant for $T\to 0$). 
A little bit away from the critical region, where the correlation
length is small compared to the system size, the susceptibility should
not show any system-size dependence, hence the $L^2$ factor must
cancel out ($\tilde{\chi}(x)\sim x^{-2}$), and we obtain
\begin{equation}
\chi_{\rm sg} \sim T^{-2\nu},
\end{equation}
at large $L$ and finite $T$. We show our result for $\chi_{\rm sg}$ in
Fig.\ \ref{figChi}. The line added corresponds to a power law with
exponent 5.0 which corresponds to $1/\nu=0.4$. The line does not fit
the data very well and there is some upward curvature of the data
which indicates a larger value for the true exponent.  This is fully
compatible with the previous results, because $1/\nu\simeq 0.295$
results in $2\nu\simeq 6.8$. To test the prediction in
Eq. \ref{eqScaleChi}, we show in Fig.\ \ref{figScaleChi} a plot of
$\chi/L^2$ as a function of $\beta L^{-1/\nu}$. We get a quality
$S=1.64$ of the finite-size scaling when considering data with $L\ge
35$ and $\chi_{\rm sg}/L^2\ge 0.3$ which means that the scaling is
good.

\begin{figure}
\begin{center}
\resizebox{0.9\linewidth}{!}{\includegraphics{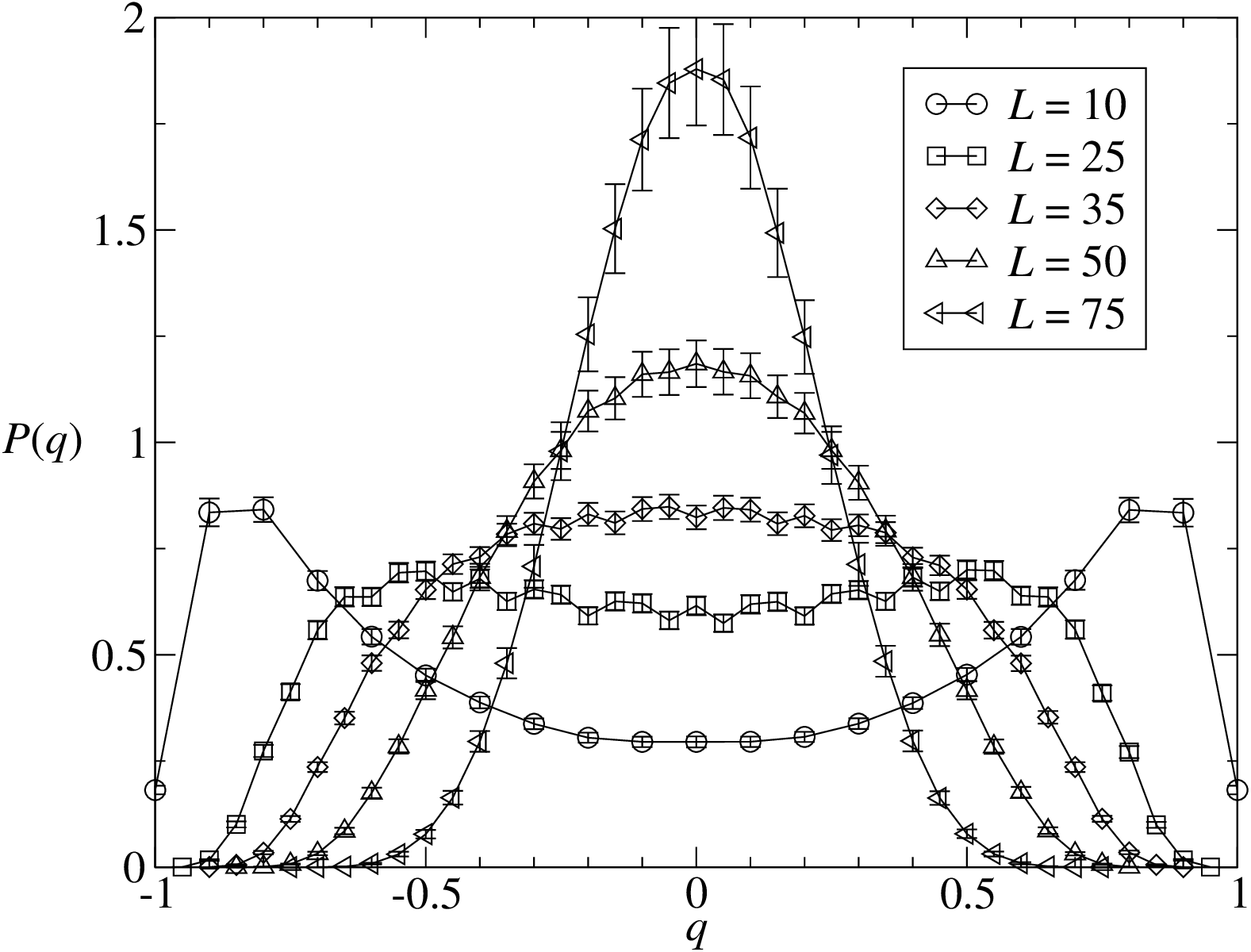}}
\end{center}
\caption{The distribution $P(q)$ of overlaps for different system
  sizes at inverse temperature $\beta=2$.}
\label{figPq2}
\end{figure}

We also studied the full distribution $P(q)$ of overlaps. The
data at $\beta=2$ is shown in Fig.\ \ref{figPq2}.  We observe that the
distributions are perfectly symmetric with respect to $q=0$, which is
another indication that our simulation is well
equilibrated. Furthermore, we can see a crossover from a two-peak
structure for small system sizes to a trivial distribution as the size
is increased.  This again is fully compatible with the notion that
there is no ordered phase a low temperatures except at $T=0$, hence
the system is paramagnetic at finite temperatures.

\begin{figure}
\begin{center}
\resizebox{0.9\linewidth}{!}{\includegraphics{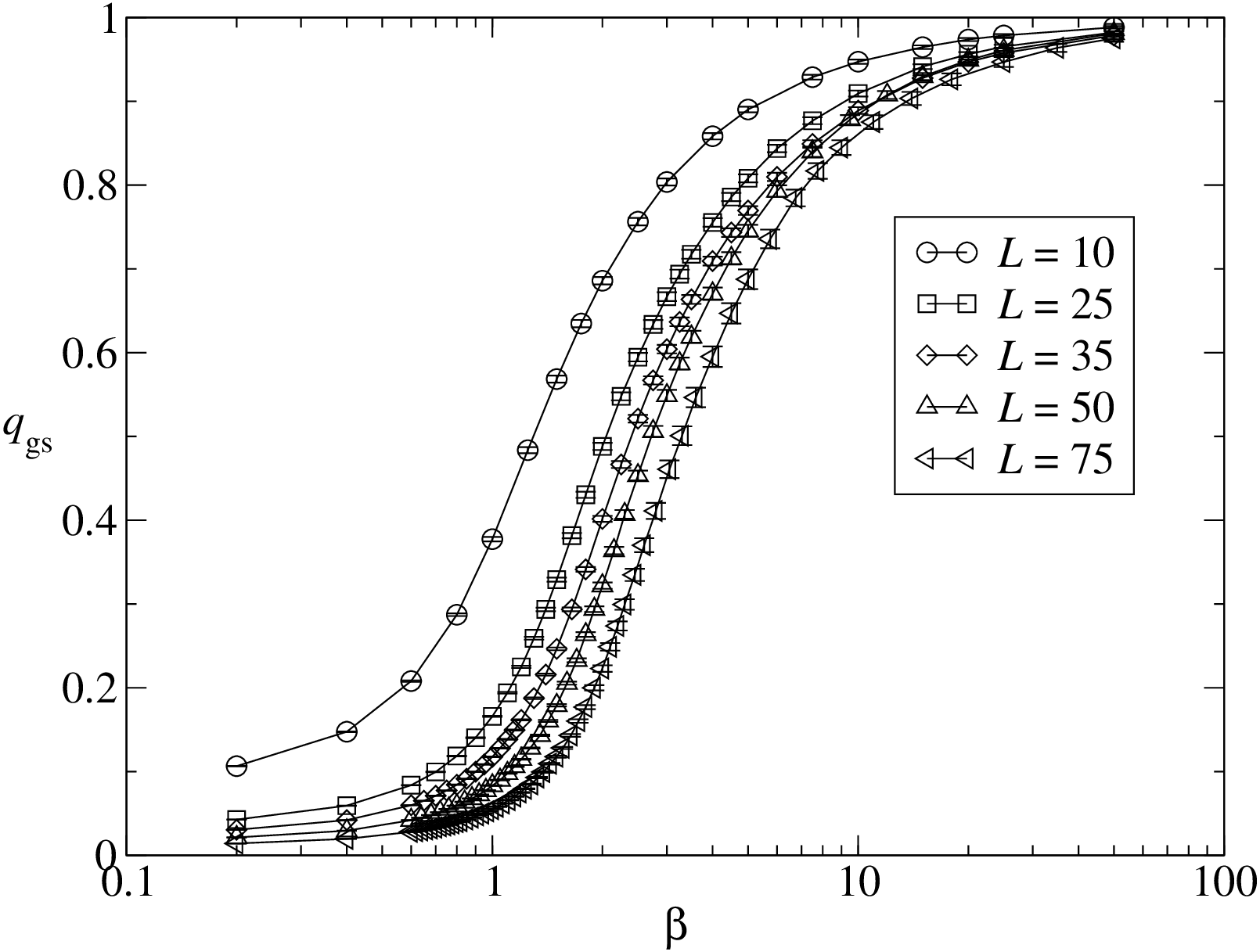}}
\end{center}
\caption{The overlap $q_{\rm gs}$ with the ground state as a function
  of the inverse temperature for different system sizes $L$.}
\label{figQgs}
\end{figure}

\begin{figure}
\begin{center}
\resizebox{0.9\linewidth}{!}{\includegraphics{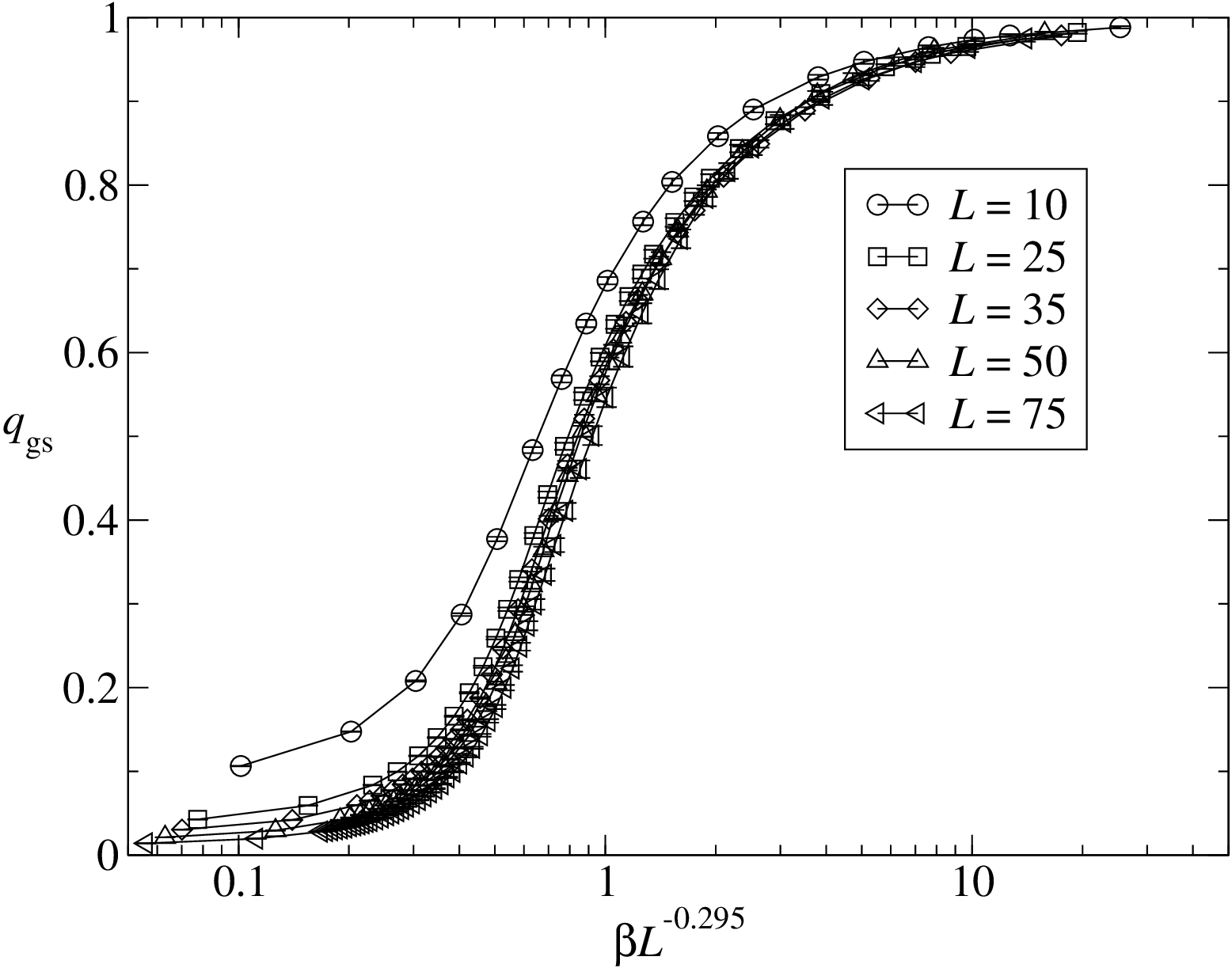}}
\end{center}
\caption{Scaling plot of the overlap with the ground state: $q_{\rm}$
  as a function of $\beta L^{-1/\nu}$, with $1/\nu=0.295$.}
\label{figScaleQgs}
\end{figure}

We furthermore looked at the overlap of the equilibrated configurations 
with the ground-state configuration \cite{rieger1996}, this can be seen
as a ``spin-glass magnetization''
\begin{equation}
q_{\rm gs} = \left[\left\langle\left|\frac1N
\sum_i S_iS_i^0\right|\right\rangle\right]_J,
\end{equation}
where $S_i^0$ denotes one of the two ground states. This is expected
\cite{rieger1996}
to scale as
\begin{equation}
q_{\rm gs} \sim \tilde{q_{\rm gs}}(L/\xi) \sim \tilde{q_{\rm gs}}(LT^\nu).
\label{eq:scale-mag}
\end{equation}
The raw data is shown in Fig.\ \ref{figQgs}, while the data rescaled
according to Eq. \ref{eq:scale-mag} with $1/\nu=0.295$ is shown in
Fig.\ \ref{figScaleQgs}. When considering data with $L\ge35$ and
$q_{\rm gs}\ge 0.5$ we find quality $S=2.50$. The quality of this
scaling is lower than the one for the Binder cumulant,
nevertheless, the result still supports the findings from
above. Interestingly, in Ref. \onlinecite{rieger1996} a similar value
$1/\nu=0.28$ was already found whereas only small systems $L\le 12$
at high temperatures ($\beta<1.4$) where studied.

\begin{figure}
\begin{center}
\resizebox{0.9\linewidth}{!}{\includegraphics{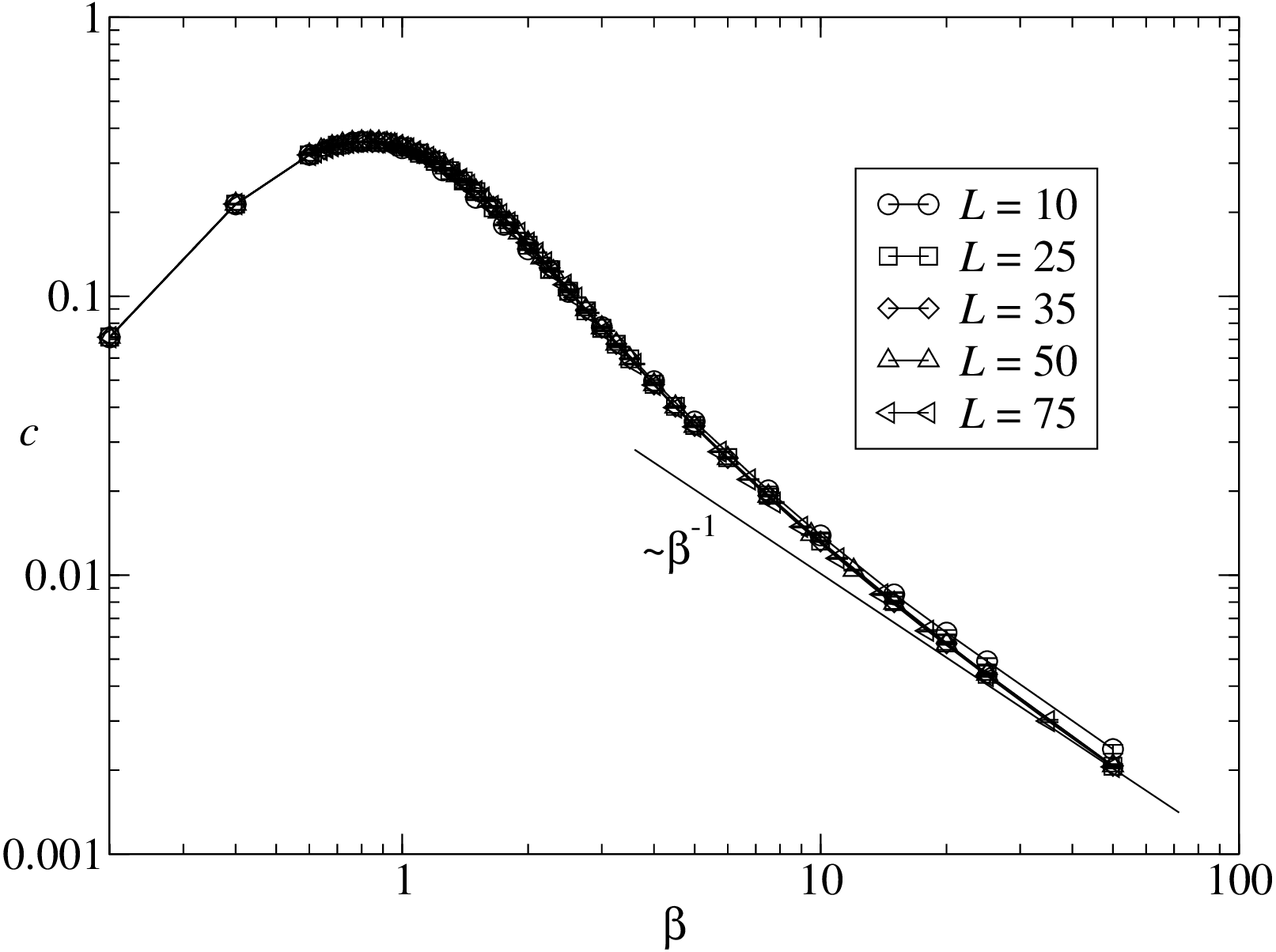}}
\end{center}
\caption{The specific heat $c$ as a function of inverse temperature
  $\beta$ for different system sizes $L$. The straight line is a
  putative asymptotic behavior $c\sim T$ when $T\to 0$.}
\label{figSpecHeat}
\end{figure}

We also studied the specific heat
\begin{equation}
c = \frac1N \left[\frac{dE}{dT}\right]_J =
\frac{\beta^2}N\left[\left\langle({\cal H}
-\langle {\cal H}\rangle)^2\right\rangle\right]_J.
\end{equation}
Although the specific heat is not expected to show any singularity for
the spin-glass transition, it is nevertheless interesting to compare
with previous studies.  The data is shown in Fig.\
\ref{figSpecHeat}. The specific heat presents a finite peak around
$\beta=0.82$ known as the ``Schottky anomaly''. The decay of the
specific heat at small $T$ goes roughly as $c\sim T$ similarly to
previous studies using transfer-matrix calculations on long stripes
\cite{cheung1983b} respectively hierarchical lattices
\cite{andrade2003}. Nevertheless the data is not perfectly described
by a power low even at such low temperatures and it is not clear what
the asymptotic behavior truely is.

Finally, we have also directly evaluated the correlation length $\xi$ by
measuring spin-spin correlations as a function of the separation of
the spins and fitting suitable functions to the data. Since the
behavior of the correlation is already studied in detail in
Ref. \onlinecite{katzgraber2004}, we do not go into details here. We only
mention that our results are fully compatible with the results of
Ref. \onlinecite{katzgraber2004}: when going to large system sizes and
studying low temperatures, we observe a behavior compatible with
$-1/\nu=-0.295$.

\section{Summary and Conclusions}

We have studied the low-temperature behavior of two-dimensional spin
glasses with a Gaussian distribution of the interactions. Using a
sophisticated cluster algorithm in combination with exploring
lowest-energy excitations close to $T=0$, we were able to equilibrate
large system sizes up to $L=75$ down to very small temperatures
$T=0.02$.

We have studied several thermodynamical quantities, like Binder
cumulant, susceptibility, distribution of overlaps, overlap with the
ground state and specific heat. Our main findings are as follows.
From the Binder cumulant, and the distribution of overlaps, we see
that no stable spin-glass phase at finite temperature exists,
i.e. $T_c=0$ in accordance to recent studies. From the scaling
behavior of the Binder cumulant and the susceptibility, we find that
the correlation length diverges algebraically for $T\to 0$, in
contrast to the model with a bimodal distribution of the interactions
\cite{houdayer2001}.  The main open question was, whether the exponent
$\theta\approx -0.29$, describing the scaling of droplet and
domain-wall excitations, is equal to $-1/\nu$, $\nu$ being the
exponent of the correlation length. Our results $-1/\nu=-0.295(30)$
obtained for large sizes and at low temperatures indeed support
$\theta=-1/\nu$.

To conclude, when taking ground-state studies into
account: the thermodynamic behavior of the two-dimensional
Gaussian Ising spin glass is trivial for finite but low temperatures
and governed by one single exponent $-1/\nu=\theta\approx -0.29$, as
predicted by the droplet picture simple renormalization arguments.

\section{Acknowledgments}
The authors thank H.G. Katzgraber,  L.W. Lee, and A.P. Young for
helpful discussions and sending us their preprint prior to submission.
 AKH obtained financial support from the
{\em VolkswagenStiftung} (Germany) within the program
``Nachwuchsgruppen an Universitäten''. This work was 
supported in part by the European Community's Human Potential
 Program under contract number HPRN-CT-2002-00307, DYGLAGEMEM.

\appendix

\section{Determining the quality of the scaling laws}
To determine the quality of the scaling laws we need some quality
criterium that somehow measures the distance of the data to the master
curve. The difficulty arise from the fact that the master curve is
unknown and must be determined from the data. Some years ago,
Kawashima and Ito \cite{kawashima1993} proposed a method. We present
here a refinement which, according to our experience,  seemed to be
 stabler and more precise.

After applying the scaling law, we have $k$ sets of points. Each set
is composed of $n_i$ points ($i=1\ldots k$) of the form
$(x_{ij},y_{ij},dy_{ij})$ with $dy$ being the standard error on $y$
and $j=1\ldots n_i$. In the following we suppose that
$x_{i1}<x_{i2}<\ldots<x_{in_i}$. We define the quality as
\begin{equation}
S = \frac1{\cal N}\sum_{i, j} \frac{(y_{ij}-Y_{ij})^2}{dy_{ij}^2+dY_{ij}^2},
\end{equation}
where $Y_{ij}$ and $dY_{ij}$ are the estimated position and standard
error of the master curve at $x_{ij}$. $\cal N$ is the number of terms
in the sum (we only consider the terms for which $Y_{ij}$ and
$dY_{ij}$ are defined).

To define $Y_{ij}$ and $dY_{ij}$, we first select a set of points as
follow: in each set $i'\ne i$, we select two points $j'$ and $j'+1$
such that $x_{i'j'}\le x_{ij}\le x_{i'(j'+1)}$, if there are no such
points in a set, we do not select any point from that set (the set
does not determine the position of the master curve for this value of
$x$). If this procedure selects no point at all then $Y_{ij}$ and
$dY_{ij}$ are undefined for point $ij$ and it does not contribute to
$S$ (this happens if set $i$ is alone in this region of $x$ and is the
master curve by itself). We now compute the linear fit through the
selected points $(x_l, y_l, dy_l)$, $l=1\ldots m$ and $Y_{ij}$ is the
value of that straight line at $x_{ij}$ and $dY_{ij}$ is the
associated standard error, namely
\begin{equation}
Y_{ij} = \frac{K_{xx}K_y-K_xK_{xy}}\Delta + x_{ij}\frac{KK_{xy}-K_xK_y}\Delta
\end{equation}
and
\begin{equation}
dY_{ij}^2 = \frac1\Delta (K_{xx}-2x_{ij}K_x+x_{ij}^2 K)
\end{equation}
with $w_l=1/dy_l^2$, $K=\sum w_l$, $K_x=\sum w_l x_l$, $K_y=\sum w_l
y_l$, $K_{xx} = \sum w_l x_l^2$, $K_{xy} = \sum w_l x_ly_l$ and
$\Delta=KK_{xx}-K_x^2$.

The quality $S$ measures the mean square distance to the master curve
of the sets in unit of standard errors. It should thus be around one
if the data really collapse to a single curve and much larger otherwise.

\end{document}